\pdfoutput=1

\documentclass[11pt]{article}

\usepackage[final]{acl}

\usepackage{times}
\usepackage{latexsym}

\usepackage[T1]{fontenc}

\usepackage[utf8]{inputenc}

\usepackage{microtype}

\usepackage{inconsolata}

\usepackage{graphicx}

\title{Interaction-Required Suggestions for Control, Ownership, and Awareness in Human-AI Co-Writing}

\author{
  Kenneth C. Arnold \\
  Calvin University \\
  \texttt{kcarnold@alum.mit.edu}
  \And
  Jiho Kim \\
  Calvin University \\
  \texttt{jihokim8@acm.org}
}

\begin{document}
\maketitle

\begin{abstract}

This paper explores interaction designs for generative AI interfaces that necessitate human involvement throughout the generation process. We argue that such interfaces can promote cognitive engagement, agency, and thoughtful decision-making. Through a case study in text revision, we present and analyze two interaction techniques: (1) using a predictive-text interaction to type the assistant's response to a revision request, and (2) highlighting potential edit opportunities in a document. Our implementations demonstrate how these approaches reveal the landscape of writing possibilities and enable fine-grained control. We discuss implications for human-AI writing partnerships and future interaction design directions.

\end{abstract}

\section{Introduction}

Current chatbot interfaces for large language models like ChatGPT, Claude, and Gemini limit interaction to a turn-taking conversation, even though the underlying models could support more versatile interactions, especially for writing tasks.

In this paper, we begin to explore the design space of interactions that people can have with model outputs, focusing on the potential opportunities presented by interactions where human initiative is \emph{required} for completing a task. Although these interactions are, by construction, less efficient at producing plausible outputs, we aim to explore the potential benefits they might offer in control, ownership, visibility of the solution space, and feedback for model tuning.

We present two interaction techniques for revision in writing: predictive-text and opportunity highlighting. The first technique adapts the familiar predictive-text interaction (top-$k$ suggestions or free typing) found on mobile devices to allow people to type the \emph{assistant's response} word by word. The second technique visualizes alternative (and sometimes divergent) choices for revising text according to a writer-specified goal.

\begin{figure}
\centering
\includegraphics[width=1.0\columnwidth]{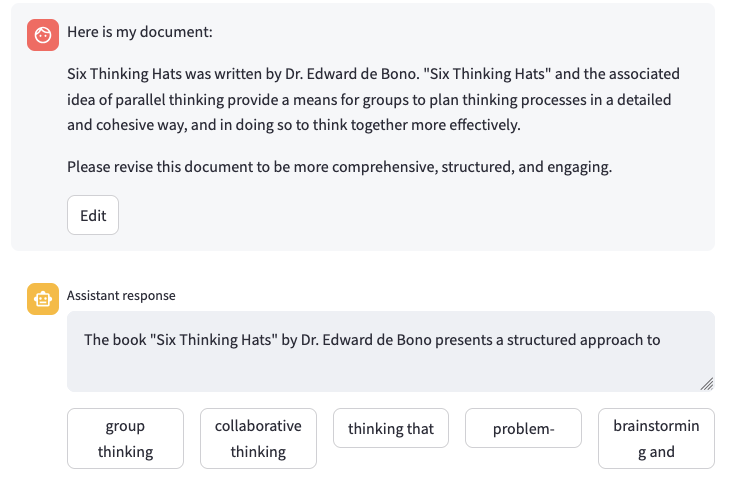}
    \caption{\label{fig:type-agent-response}Predictive text interaction repurposed to type the  assistant's response}
\end{figure}

\begin{figure}
\centering
\includegraphics[width=1.0\columnwidth]{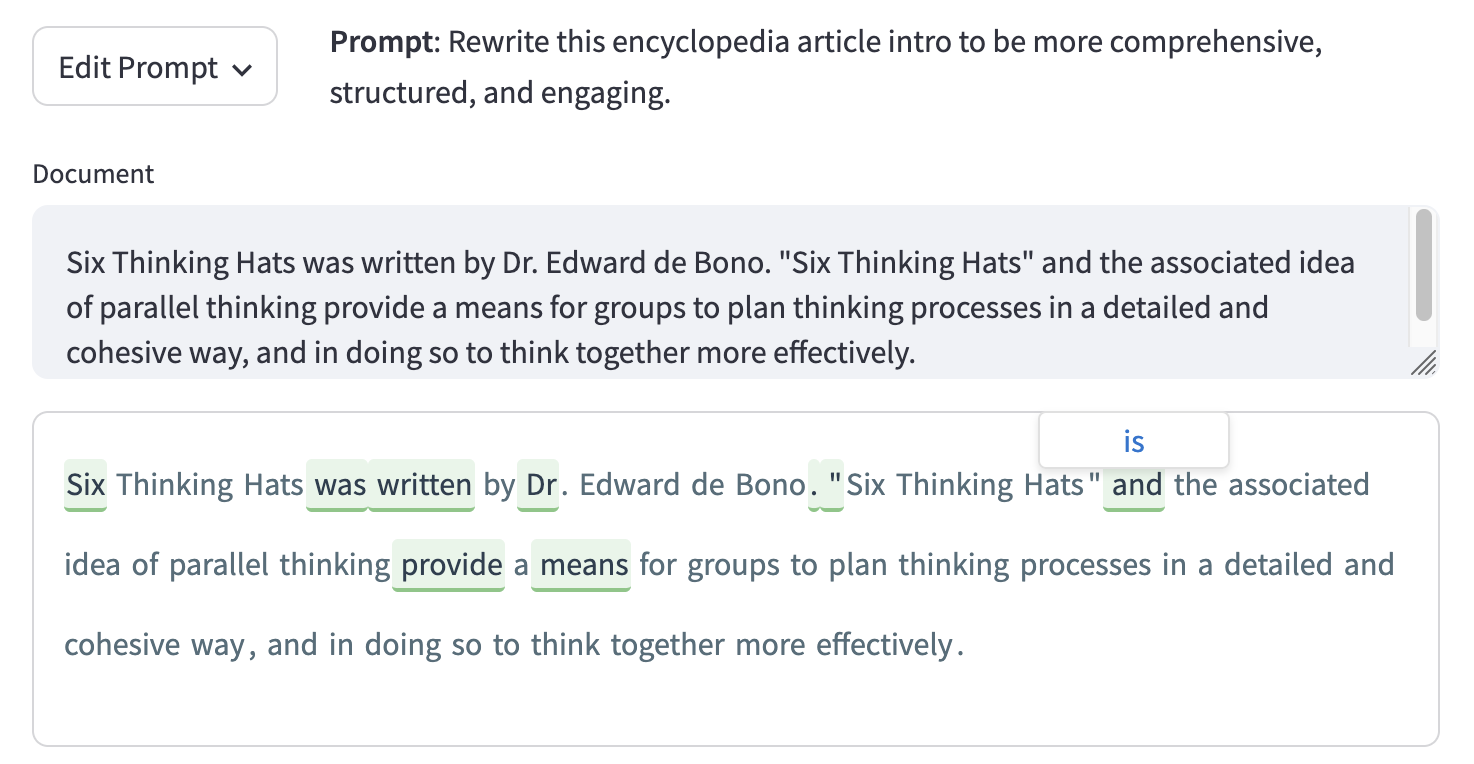}
    \caption{\label{fig:highlight-opportunities}Highlighting opportunities for divergent choices}
\end{figure}

\section{Design Principles}
\label{sec:design-principles}

AI support for writing has evolved primarily along two interaction paradigms: conversational exchanges with an assistant (as in modern chatbots) and editorial feedback systems (like inline markup in Grammarly or reflection tools like Impressona~\cite{benharrakWriterDefinedAIPersonas2024b} and Textfocals~\cite{kim2024FullAuthorshipAI}. While these paradigms have proven useful, they both place the AI in a position of either generating content or evaluating it, with humans primarily reacting to AI output.

We propose interaction-required approaches that fundamentally shift this dynamic by necessitating ongoing human involvement throughout the generation process. Our approach is guided by three design principles that emphasize cognitive partnership between humans and AI systems:

\paragraph{Prioritize cognitive engagement over efficiency} 
Although AI assistance can speed task completion, using it without cognitive engagement can lead to overconfidence~\cite{fernandes2025PerformanceMetacognitionDisconnect}, errors~\cite{dakhelGitHubCopilotAI2023}, and skill stagnation~\cite{gajos2022PeopleEngageCognitively}. 
Interactions can instead be designed to encourage writers' thoughtful participation rather than optimizing solely for speed or ease. This principle addresses how AI systems can support authentic self-expression, ownership, and accountability in writing, which many writers desire~\cite{biermann2022ToolCompanionStorywritersa,hwangItWas802024}. The literature on explainable AI systems for decision-making suggests cognitive engagement as valuable goal~\cite{dattaWhosThinkingPush2023}.

\paragraph{Enable granular control}
Rather than offering only coarse accept/reject options for completed AI outputs, interfaces could instead allow writers to influence the progress of generation. Granularity could enable just-in-time feedback that shapes the direction of AI assistance, providing a way for users to clarify their goals without having to engage in prompt refinement or writing examples.

\paragraph{Reveal the landscape of possibilities}
Interactions should make visible the alternatives available at each decision point, helping writers understand the range of options and make more informed choices. 
Prior work has explored contextual suggestions of alternative words or phrases at targeted points~(e.g., \citet{rezaABScribeRapidExploration2023,Gero2019StylisticThesaurus}), but some authors have explored interfaces for navigating through the tree of suggestions in a narrative generation context~\cite{reynolds2021MultiversalViewsLanguage}.



\subsection{Interaction-Required Suggestions}

The degree to which a writing support interface \emph{requires} interaction can be measured, in principle, by an \emph{amplification ratio}, the ratio of the entropy of system output (new text or edits) to the entropy of user input. For example, asking a chatbot to write a complete essay or make overall edits has a high amplification ratio since the input entropy is confined to the prompt. Accepting grammar suggestions also has high amplification ratio, since the user often only needs to click "Accept". 

We conjecture that LLM-powered interfaces with a low amplification ratio can be designed according to these design principles to assist writers at various points in the writing process.

\section{A Case Study in Revision}

We will present two interaction designs that embody these design principles for the purpose of revision. As a running hypothetical example, suppose Alex is a Wikipedia editor who wants to revise the introduction section for the article on "Six Thinking Hats", as it was on 2025-02-25:

\begin{quote}
    
``Six Thinking Hats was written by Dr. Edward de Bono. "Six Thinking Hats" and the associated idea of parallel thinking provide a means for groups to plan thinking processes in a detailed and cohesive way, and in doing so to think together more effectively.''
\end{quote}

We will use a revision instruction generated by Claude.ai: ``Rewrite this document to be more comprehensive, structured, and engaging.''

\subsection{Typing the Assistant's Response with Predictive Text}
\label{sec:type-agent-response}

Alex starts a chatbot conversation in the now-customary way, asking for a revision according to her goals~\ref{fig:type-agent-response}. She now sees the assistant's response being formed---but instead of seeing the assistant type its response, Alex sees an editable text box, which starts empty except for the now-familiar buttons of predictive text.

Alex starts by ignoring the prediction buttons because she realizes it would be clearer to start with ``The book'', so she starts by typing that phrase. Afterwards the predictions give the title of the book, followed by the author, which Alex readily accepts with a few taps. After that, the top 3 suggestions are ``revolutionized'', ``presents'', and ``is''; she take ``presents'', an active verb that avoids exaggeration. The next suggestions are ``a revolutionary'', ``an innovative'', and ``a groundbreaking'', which exhibit the same problem of exaggeration as before. These suggestions were probably due to Alex's prompt of ``engaging'', but the vacuous exaggeration of the suggestions indicates to Alex that she needs to consider what exactly \emph{should be} engaging about this introduction. So she pauses to read the rest of the article and concludes that the most important aspect is that the book provides a structured approach to thinking in individual and group settings. She needed to type ``a structured'', but then the predictions offered acceptable remaining words with only a bit of guidance: ``approach to thinking, both individually and collectively.''

\paragraph{Takeaways}

This interaction leverages the familiarity of the predictive-text interaction that is ubiquitous on smartphones, but the simple extension of this familiar interface to the context of typing the assistant's response to a revision request yields several unique kinds of uses:

\begin{itemize}
\item The system sometimes helps with routine tasks, like typing a book name (functioning like an adaptive copy-and-paste).
\item The same interaction can suggest alternative wordings for phrases, using the natural 3-- or 5--option button interface.
\item Unlike a chat interface, the writer can exert granular, just-in-time control over the system.
\item Some suggestions can even be provocative, leading the writer to pause and think more about what they wanted to say.
\end{itemize}

The prototype shows short phrases in prediction buttons, inspired by~\citet{Arnold2016-phrase}; next-phrase suggestions can shape writer thinking more than individual words even when not used directly~\cite{bhatInteractingNextPhraseSuggestions2023,Arnold2018:sentiment-bias,jakeschCoWritingOpinionatedLanguage2023}.

\subsection{Highlighting Edit Opportunities}

Figure~\ref{fig:highlight-opportunities} shows a different interface with the same source text and prompt. This interface shows Alex's document with highlights in places where Alex might consider making edits to enact the revision goal that she has just specified. Hovering over an opportunity highlight shows a provocative clue of what an edit there might look like. Alex notices that ``and'' is highlighted; reading the phrase she notices that the phrase (``and the associated idea of parallel thinking'') is not well connected to the main thought of the paragraph and decides to seek an alternative. Hovering the ``and'' reveals ``is'', suggesting that the next phrase could simply describe the book itself more (e.g., ``is a guide for\ldots{}'') or perhaps state something concrete about its impact (e.g., ``is the top-cited book on\ldots''). Reading the rest of the paragraph and article, Alex decides to go with the description strategy, but chooses a different word: ``describes a process for groups to plan thinking\ldots''. She makes this edit in the document and the opportunity highlights update to suggest other potential edits. She notices that the word ``detailed'' doesn't quite fit with how she understands the book; even though it is not highlighted, she hovers over it and sees an alternative, ``structured'', which seems more accurate.

\paragraph{Takeaways} 

\begin{itemize}
\item Alex retained full control over their document; all of the words are her own.
\item In contrast to editing systems like Grammarly, Alex also had detailed control (via the prompt) over what sort of edit opportunities they wanted to see.
\item The interaction allowed Alex to explore alternative choices: every word offered an alternative, even those not highlighted. 
\item The words shown in edit opportunities were sometimes substitutions but often instead offered a different semantic or grammatical direction that could be taken.
\item It is still possible for the result to be entirely AI-generated text, but that would require the writer iteratively inspecting and applying every suggested change.
\end{itemize}

\section{Discussion}

So far these interaction designs have only been evaluated informally; empirical studies with writers are needed to determine how interaction-required suggestion interfaces affect writers' sense of ownership, control, and awareness of alternatives.
Anecdotally, however (from use by the authors and a few others), both have been useful in low-level editing (trimming and clarifying wording), the predictive-text interface has been helpful for initial drafting (e.g., based on an outline), but neither are useful for larger-scale revision because they focus attention on localized choices; other tools are needed to address those needs (e.g., \citet{dangTextGenerationSupporting2022,benharrakWriterDefinedAIPersonas2024b,kim2024FullAuthorshipAI}).

Although we described a case study in revision, predictive text could be used in any assistant response. We are particularly curious about how it might have different effects across different types of tasks: open-ended tasks such as ideation, analytical tasks such as review generation, and close-ended tasks such as refactoring code.

The straightforward application of predictive text to typing the assistant's response, as we propose in Section~\ref{sec:type-agent-response}, presents opportunities to increase cognitive engagement and control over the status quo of accepting complete generated responses. Yet it is still possible to use the chatbot's words uncritically by accepting suggestions rapidly. (Should the interface be designed to allow larger-block acceptance?) And even cognitive engagement with the suggestions could still lead to a reduced sense of ownership over the result~\cite{lehmann2022SuggestionListsVs} and influence on human opinions~\cite{Arnold2018:sentiment-bias,jakeschCoWritingOpinionatedLanguage2023}. Additional exploration of the interaction design (e.g., how alternatives are visualized and navigated) is needed.

The additional control afforded by predictive text (effectively prefilling the assistant's response) affords some additional risks for users to jailbreak the model~\cite{andriushchenko2024JailbreakingLeadingSafetyAligned}. However, since prefilling is part of many commercial LLM APIs, we doubt that this interaction design presents significant marginal risk.

Predictive text can be viewed as an interactive visualization of high-probability local alternatives within a sequence of categorical choices (e.g., Figure~\ref{fig:type-agent-response} shows two-token predictions to provide awareness of where each suggestion could be going.\footnote{We plan to implement the phrase preview interaction of \citet{Arnold2016-phrase} to enable writers to see larger phrases without having to use all of them.} From this perspective a wide range of interactive visualization techniques are possible, such as the Dasher text entry system~\cite{ward2000DasherDataEntry} (which may have accessibility benefits as well).
Design dimensions of these visualizations include the granularity of suggestions (words, phrases, or larger units such as copy-pasted text from a writer's other drafts) and how interacting with the suggestion affects the surrounding text. The effects of these design decisions might vary by stage of the writing process.

The opportunity highlighting interface explored an extreme design position of being minimally prescriptive in AI help, but relaxing that extreme could yield a range of alternative interaction designs. For example, it could incorporate an interactive visualization where the writer could navigate through contextual alternatives at any point.

Both of these models presuppose autoregressive (left-to-right) language modeling, but additional types of interaction might be enabled by emerging model types based on out-of-order modeling or diffusion LLMs~\cite{sahoo2024SimpleEffectiveMasked}.

Although prior work has explored the effects of generating different kinds of content with LLMs on writer reactions \cite{benharrakWriterDefinedAIPersonas2024b, kim2024FullAuthorshipAI, zhou2024AiLludeInvestigating}, this work keeps the task for the LLM unchanged and explores the kinds of interactions that people can have with the inference process.

Interaction-required suggestions are a source of rich feedback data for reward-based language model training and personalization. Unlike static documents, the interaction logs with a conversational predictive text system would include what suggestions were made but not taken, providing a fine-grained human feedback signal. These feedback signals can be used for updating a language model~\cite{wu2023FineGrainedHumanFeedback,arnold2017CounterfactualLanguageModel}.

\paragraph{Conclusion}
With continuously increasing capabilities of LLMs, the difference between augmenting and replacing human thinking is a question not of system capabilities but of interaction design. The interaction-required approaches we've presented demonstrate how small shifts in interface design can fundamentally change the nature of human-AI partnership in writing. By prioritizing cognitive engagement, enabling granular control, and revealing the landscape of possibilities, we can design AI writing interfaces that help us think not \emph{less} but \emph{better}—maintaining human agency while still benefiting from AI capabilities.

\section*{Acknowledgments}

We thank Hannah Yoo, Jason Chew, Juyeong Kim, Heonjae Kwon, Ray Flanagan, and the anonymous reviewers for their input and feedback. This work is supported by NSF CRII award 224614.

\bibliography{custom}

\section*{Appendix}

\subsection*{Implementation Details}

The prototypes described here were implemented using a Streamlit frontend and a backend using the Hugging Face Transformers library~\cite{HuggingFaceTransformers}. Full source code and demo is available at \url{https://huggingface.co/spaces/CalvinU/writing-prototypes}.

Both of these systems rely on language model functionality that is not typically exposed in efficient ways in commercial APIs\footnote{For example, prompt logprobs, needed for highlighting, was part of the OpenAI text completions API but was never added to the chat completions API}, but are straightforward to implement when given direct access to the forward pass of the model, which computes next-token distributions for all tokens in the context (including both ``user'' and ``assistant'' messages). The implementation in our demo uses the Gemma 2 9B model released by Google~\cite{team2024Gemma2Improving}.

The predictive text interface first computes the top-$k$ (e.g., 3 or 5) next tokens, then constructs a short phrase (in the demo, a single additional token) by greedy generation from each of those options. With careful management of the key-value cache, this generation readily completes at interactive speed on commodity hardware. Predictive-text coding systems like GitHub Copilot served as informal prototypes of this interaction (since instructions can be entered as code comments), but they did not reveal the landscape of possibilities (see section~\ref{sec:design-principles} on Design Principles) in the way that smartphone keyboards and our system do.

The highlighting interface constructs a pseudo-conversation by where the user message is the revision prompt concatenated with the original document and the assistant message is the original document repeated unchanged. Rather than generate additional tokens, we simply compute the next-token distributions for all tokens in the ``assistant'' message corresponding to the user's document. The frontend highlights the tokens where the model gives a higher score to a token other than the one in the original document. Mouseover hovers show an alternative token; for tokens where the argmax prediction matched the original document (which are typically the majority of tokens), the hover shows the 2nd highest-scored option.

\end{document}